\documentclass{mem}
\usepackage{natbib}\usepackage{txfonts}\usepackage{balance}
\usepackage{graphicx}
\usepackage[a4paper]{hyperref}
\idline{75}{282}
\begin{document}
\def\teff{$T\rm_{eff }$}
\def\kms{$\mathrm {km s}^{-1}$}

\title{
Constraining cosmic reionization with quasar, gamma ray burst, and Ly$\alpha$ emitter observations
}

   \subtitle{}

\author{
S. \, Gallerani\inst{1}, 
A. \, Ferrara\inst{2},
T. Roy \,Choudhury\inst{3},
X. \,Fan\inst{4},
R. \,Salvaterra\inst{5},
P. \,Dayal\inst{6},
          }

  \offprints{S. Gallerani}

\institute{
INAF-Osservatorio Astronomico di Roma,
via di Frascati 33, 00040 Monte Porzio Catone, Italy
(\email{gallerani@oa-roma.inaf.it})
\and
Scuola Normale Superiore,
Piazza dei Cavalieri 7, 
56126 Pisa, Italy
\and
Harish-Chandra Research Institute, 
Chhatnag Road, 
Jhusi, 
Allahabad 211 019, India
\and
Steward Observatory, 933 N. Cherry Ave, Tucson, AZ 85721-0065, US
\and
INAF-Osservatorio Astronomico di Brera, via E. Bianchi 46, 23807 Merate (LC), Italy
\and
Scuola Internazionale Superiore di Studi Avanzati, via Beirut 2-4, 34151 Trieste, Italy
}

\authorrunning{S. Gallerani et al.}

\titlerunning{Constraining reionization with QSO, GRB, and LAE observations}

\abstract{
We investigate the cosmic reionization history by comparing semi-analytical models of the Ly$\alpha$ forest with observations of high-$z$ quasars and gamma ray bursts absorption spectra. In order to constrain the reionization epoch $z_{\rm rei}$, we consider two physically motivated scenarios in which reionization ends either early (ERM, $z_{\rm rei}\gtrsim 7$) or late (LRM, $z_{\rm rei}\approx 6$).
We analyze the transmitted flux in a sample of 17 QSOs spectra at $5.7\leq z_{\rm em}\leq 6.4$ and in the spectrum of the GRB 050904 at $z=6.3$, studying the wide dark portions (gaps) in the observed absorption spectra. By comparing the statistics of these spectral features with our models, we conclude that current observational data do not require any sudden change in the ionization state of the IGM at $z\approx 6$, favouring indeed a highly ionized Universe at these epochs, as predicted by the ERM. Moreover, we test the predictions of this model through Ly$\alpha$ emitters observations, finding that the ERM provide a good fit to the evolution of the luminosity function of Ly$\alpha$ emitting galaxies in the redshift range $z=5.7-6.5$. The overall result points towards an extended 
reionization process which starts at $z\gtrsim 11$ and completes at $z_{\rm rei}\gtrsim 7$, in agreement with the recent WMAP5 
data.   
}
\maketitle{}

\section{Introduction}

In the last few years a possible tension has been identified between WMAP5 data \citep{dunkley} and SDSS observations of quasar (QSO) absorption 
spectra \citep{fan}, the former being consistent with an epoch of reionization 
$z_{\rm rei}\approx 11$, the latter suggesting $z_{\rm rei}\approx 6$. Long Gamma Ray 
Bursts (GRB) may constitute a complementary way to study the reionization 
process, possibly probing $z>6$ \citep{taglia,greiner,salva}. Moreover, an increasing number of Lyman Alpha 
Emitters (LAE) are routinely found at $z>6$ \citep{stark}.\\ 
The ultraviolet radiation emitted by a QSO/GRB can suffer resonant Ly$\alpha$ scattering as it propagates
through the intergalactic neutral hydrogen. In this process, photons are removed from the line of
sight (LOS)  resulting in an attenuation of the source flux, the so-called Gunn-Peterson (GP) 
effect. For these reason QSO/GRB absorption spectra are recognized as very powerful tools for measuring the neutral hydrogen fraction $x_{\rm HI}$ in the InterGalactic Medium (IGM), and hence for determining $z_{\rm rei}$. Moreover, the observed Ly$\alpha$ Luminosity Function (LF) of LAEs can be used to infer the ionization state of the IGM at redshifts
close to those of the star-forming galaxies and hence to reconstruct the reionization history. In this work, we compare the predictions of theoretical models of reionization with QSOs, GRBs, and LAEs observations.   
\section{Reionization models}
\begin{figure}[t!]
\resizebox{\hsize}{!}{\includegraphics[clip=true]{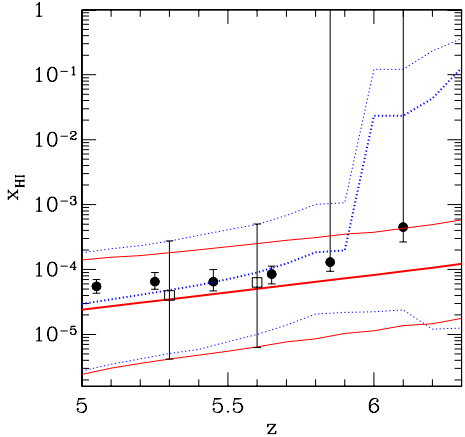}}
\caption{\footnotesize
 Evolution of the neutral hydrogen fraction. 
Thick red solid (blue dotted) lines represent average results over 100 LOS for the ERM (LRM), while the thin lines 
denote the upper and lower neutral hydrogen fraction extremes in each redshift 
interval. Solid circles represent $x_{\rm HI}$ estimates by \citet{fan}; 
empty squares denote the results obtained in this work.
}
\label{eta}
\end{figure}
We present a semi-analytical model of cosmic reionization, developed by \citet{cf05} and \citet{cf06} and further refined by \citet{cf08}. This model, hereafter called CF05, allows to build a wide range of reionization scenarios which match many observational constraints (ranging from WMAP5 to SDSS data) and differ for $z_{\rm rei}$ and for the IGM properties at $z\geq 6$. In order to constrain reionization we have adopted the CF05 model to set up two different reionization scenarios, namely: (i) an early reionization model (ERM), favored by WMAP data, characterized by a highly ionized IGM at $z\approx 6$, and (ii) a late reionization model (LRM), in which $z_{\rm rei}\approx 6$, as suggested by the SDSS results. The $x_{\rm HI}$ predicted by the two models is shown in Fig.1. Moreover, we have developed a semi-analytical model to simulate QSO absorption spectra starting from the CF05 predictions \citep{g06}.  

\section{RESULTS}
\subsection{QSOs absorption spectra}
\begin{figure*}[]
\resizebox{\hsize}{!}{\includegraphics[clip=true]{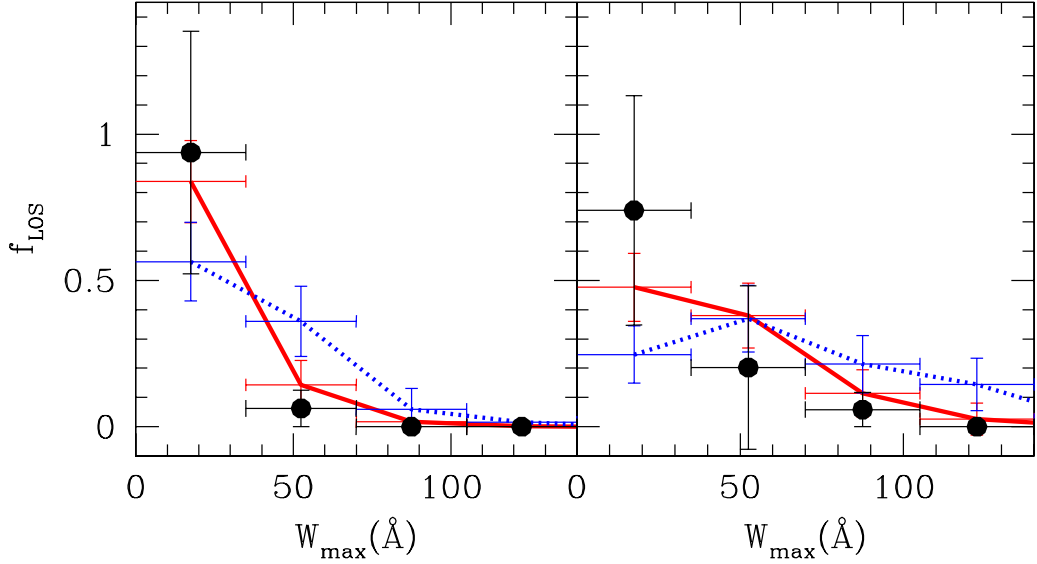}}
\caption{
\footnotesize
LGW distribution for QSOs with $z_{\rm em}<6$ (left) and $z_{\rm em}>6$ (right). 
Filled circles represent observational data. Solid red (dotted blue) lines show the ERM 
(LRM) predictions. Vertical error bars measure poissonian noise, horizontal errors define 
the bin for the gap widths.
}
\label{li_vhel}
\end{figure*}
We use observational data including 17 QSOs obtained by \citet{fan}.
We divide the observed spectra into two redshift-selected sub-samples: 
the ``Low-Redshift'' (LR) sample ($5.7 < z_{\rm em} < 6$), and
the ``High-Redshift'' (HR) one ($ 6 < z_{\rm em} < 6.4$). 
By comparing the the Largest Gap\footnote{Gaps are defined as contiguous regions of the spectrum characterized by a transmitted flux lower than a flux treshold $F_{\rm th}=0.1$.} Width (LGW) distribution\footnote{The LGW distribution quantifies the fraction of LOS which are characterized by the largest gap of a given width.} obtained from the observed spectra with simulations (Fig.2) we find that both the ERM and the LRM provide a good fit to 
observational data. We exploit this agreement to derive an estimate of $x_{\rm HI}$. We find $\log_{10}x_{\rm HI}=-4.4^{+0.84}_{-0.90}$ at $z\approx 5.3$, and $\log_{10}x_{\rm HI}=-4.2^{+0.84}_{-1.0}$ at $z\approx 5.6$.
Although the predicted LGW distributions are quite similar for the two models 
considered, in the HR case we 
find that a neutral hydrogen fraction at $z\approx 6$ 
higher than that one predicted by the LRM would imply an even worst 
agreement with observations, since a more abundant HI would produce a lower 
(higher) fraction of LOS characterized by the largest gap smaller (higher) 
than $40$ \AA\ with respect to observations. 
Thus, this study suggests $x_{\rm HI}<0.36$ at $z=6.32$ \citep{g08a}.
\subsection{GRBs absorption spectra}
\begin{figure*}[]
\resizebox{\hsize}{!}{\includegraphics[clip=true]{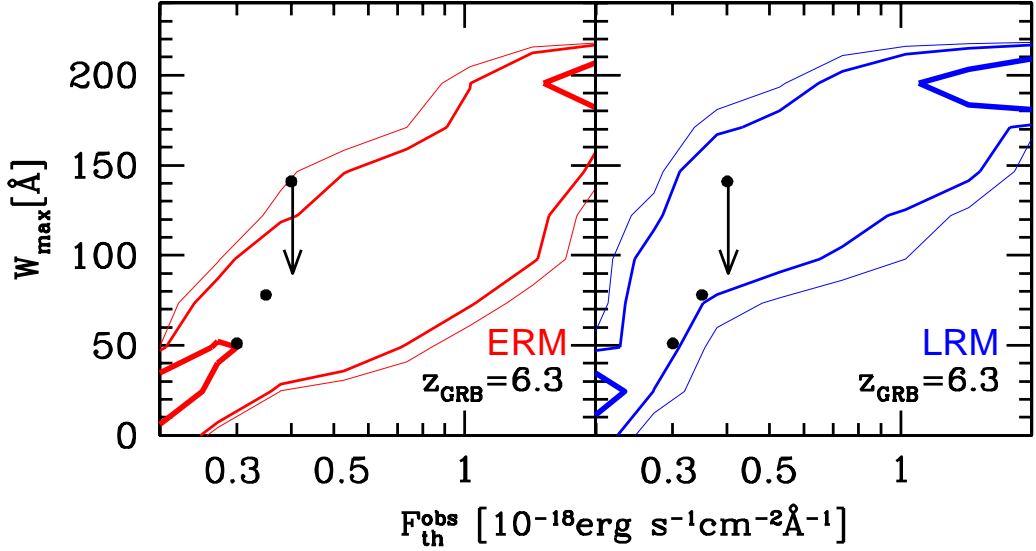}}
\caption{
\footnotesize
Isocontours of the probability that the afterglow spectrum 
associated with a GRB at redshift $z_{\rm GRB}=6.3$, contains a 
largest 
gap of size in the range $[W_{\rm max}, W_{\rm max}+dW]$, with $dW=20$~\AA, for a flux threshold $F_{\rm th}$.  
The left (right) panel shows the results for the ERM (LRM). The isocontours correspond to probability of 5\%, 10\%, and 40\%. The black points indicates the position in the $(W_{\rm max}, F_{\rm th}^{\rm obs})$ plane of 
GRB 050904. The point with arrow means that the gap size should be considered as an upper limit, since the corresponding dark region could be affected by the presence of a DLA.
}
\label{li_vhel}
\end{figure*}
We have analyzed the optical afterglow spectrum of the GRB 0509004 at $z=6.29$ by measuring the LGWs in its Ly$\alpha$ forest as a function of the flux threshold $F_{\rm th}$ used to define gaps. Starting from a large sample of synthetic spectra, we have computed the probability to find gaps of these widths in absorption spectra obtained through the ERM and LRM (Fig.3). We find that the ERM is two times more predictable than the LRM, thus confirming the results found in the case of QSOs, i.e. that a highly ionized Universe at $z\approx 6$ is favored by current observations \citep{g08b}.

\subsection{LAEs luminosity function}
We have built up a semi-analytic model of LAEs to compute the evolution of the $Ly\alpha$ LF of star forming galaxies at redshifts approaching $z_{\rm rei}$. We have fixed $x_{\rm HI}$ to the predictions of the ERM and LRM at $z=5.7$ and $z=6.6$ leaving only the star formation efficiency and the effective escape fraction of  photons as free parameters. The results are shown in Fig.4. The ERM provides a good fit to observational data, implying that no sudden change in the ionization state of the IGM at $z\approx 6$ is required to explain the evolution of the $Ly\alpha$ LF \citep{d08}. The ERM has been also used by \citet{d09a} to predict the LAEs LF up to $z=7.6$ and by \citet{d09b} to study the dust content of LAEs. 
\begin{figure}[]
\resizebox{\hsize}{!}{\includegraphics[clip=true]{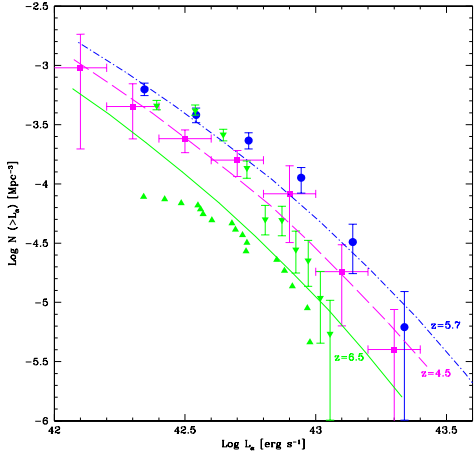}}
\caption{
\footnotesize
LAE LF for the ERM at $z=5.7$ (dot-dashed), $z=6.56$ (solid). Points represent the data at $z=5.7$ (circles; \citet{shima}), $z=6.56$ (downward/upward triangles represent upper/lower limits; \citet{kashi}).
}
\label{li_vhel}
\end{figure}

\section{Conclusions}
We have presented a semi-analytical model which allows to simulate QSOs, GRBs, and LAEs spectra. 
Synthetic spectra have been analyzed statistically and the theoretical predictions compared with observations. We have shown that current data do not require any sudden change in the IGM ionization level at $z\approx 6$, favouring a highly ionized IGM at these epochs. The overall result points towards an extended 
reionization process which starts at $z\gtrsim 11$ and completes at $z_{\rm rei}\gtrsim 7$, in agreement with the recent WMAP5 
data. 

\bibliographystyle{aa}

\end{document}